\documentclass[aps,final]{revtex4-1}  

\usepackage{amsfonts}
\usepackage{amsmath}
\usepackage{amssymb}
\usepackage{graphicx}
\usepackage{color}

\def\prod{\mathrm{prod}}
\def\hil{{\mathcal H}}
\def\kil{{\mathcal K}}

\def\C{{\mathcal C}}
\def\D{{\mathcal D}}
\def\M{{\mathcal M}}
\def\X{{\mathcal X}}
\def\S{{\mathcal D}}
\def\ep{\varepsilon}
\def\bz{\left(}
\def\jz{\right)}
\def\half{\frac{1}{2}}
\def\vfi{\varphi}

\newtheorem{theorem}{Theorem}

\newtheorem{corollary}[theorem]{Corollary}

\newtheorem{definition}[theorem]{Definition}

\newtheorem{lemma}[theorem]{Lemma}

\newtheorem{proposition}[theorem]{Proposition}
\newtheorem{remark}[theorem]{Remark}

\newenvironment{proof}[1][Proof]{\noindent\textbf{#1.} }{\hfill \rule{0.6em}{0.6em}\\}

 \DeclareMathOperator{\tr}{Tr}

\newcommand{\ket}[1]{| #1 \rangle}

\newcommand{\B}{{\tilde{B}}}

\newcommand{\osigma}{{\overline{\sigma}}}

\newcommand{\be}{{\mathbf e}}

        \def\cB{{\cal B}}
\def\cC{{\cal C}}
\def\cD{{\cal D}}        

        \def\cH{{\cal H}}

\def\cX{{\cal X}}

\def\0{{\mathbf{0}}}
\def\1{{\mathbf{1}}}
\def\2{{\mathbf{2}}}
\def\3{{\mathbf{3}}}
\def\4{{\mathbf{4}}}
\def\5{{\mathbf{5}}}
\def\6{{\mathbf{6}}}

\def\7{{\mathbf{7}}}
\def\8{{\mathbf{8}}}
\def\9{{\mathbf{9}}}

\def\bbN{{\mathbb{N}}}

\def\be{\begin{equation}}
\def\ee{\end{equation}}
\def\bea{\begin{eqnarray}}
\def\eea{\end{eqnarray}}
\def\reff#1{(\ref{#1})}
\def\eps{\varepsilon}

\def\hil{{\mathcal H}}
\def\kil{{\mathcal K}}
\def\B{{\mathcal B}}
\def\C{{\mathcal C}}
\def\D{{\mathcal D}}
\def\M{{\mathcal M}}
\def\X{{\mathcal X}}
\def\S{{\mathcal D}}
\def\ep{\varepsilon}
\def\bz{\left(}
\def\jz{\right)}
\def\half{\frac{1}{2}}
\def\vfi{\varphi}

\def\n{^{(n)}}

\def\bN{\mathbb{N}}
\def\d{d_{\mathrm{op}}}
\def\inv{^{-1}}

\newcommand{\norm}[1]{\left\| #1\right\|}
\newcommand{\sr}[2]{D\bz #1 \| #2\jz}
\newcommand{\rsr}[3]{D_{#3}\bz #1 \| #2\jz}
\newcommand{\srmax}[2]{D_{\mathrm{max}}\bz #1 \| #2\jz}
\newcommand{\srmaxs}[3]{D_{\mathrm{max}}^{#3}\bz #1 \| #2\jz}

\newcommand{\ds}{\mbox{ }\mbox{ }}

\newcommand{\diad}[2]{|#1\rangle\langle #2|}
\newcommand{\pr}[1]{\diad{#1}{#1}}

\newcommand{\inner}[2]{\langle #1,#2\rangle}
\newcommand{\ki}[1]{\textit{#1}}
\newcommand{\kiii}[1]{\textit{#1}}

\DeclareMathOperator{\Tr}{Tr}

\DeclareMathOperator{\supp}{supp}

\def\P{{\mathcal P}}

\begin{document}

\title{A smooth entropy approach to quantum hypothesis testing and the classical capacity of quantum channels
}
 \author{Nilanjana Datta}
\affiliation{Statistical Laboratory, University of Cambridge, Wilberforce Road, Cambridge CB3 0WB, United Kingdom}
 \author{Mil\'an Mosonyi} 
 \affiliation{School of Mathematics, University of Bristol, University Walk, Bristol, BS8 1TW, UK\\
 and\\
 Department of Analysis, Budapest University of Technology and Economics,
 Egry J\'ozsef u.~1., Budapest, 1111 Hungary}
 \author{Min-Hsiu Hsieh}
 \affiliation{Centre for Quantum Computation \& Intelligent Systems (QCIS),
 Faculty of Engineering and Information Technology (FEIT)
 University of Technology Sydney (UTS), NSW 2007, Australia}
 \author{ Fernando G.S.L.~Brand\~ao}
\affiliation{Institute for Theoretical Physics, ETH Z\"urich, 8093 Zurich, Switzerland}


\begin{abstract}
We use the smooth entropy approach to treat the problems of binary quantum hypothesis testing and the transmission of classical information through a quantum channel. We provide lower and upper bounds on the optimal type II error of quantum hypothesis testing in terms of the smooth max-relative entropy of the two states representing the two hypotheses. 
Using then a relative entropy version of the Quantum Asymptotic Equipartition Property (QAEP), we can recover the strong converse rate of the i.i.d.~hypothesis testing problem in the asymptotics. On the other hand, combining Stein's lemma with our bounds, we obtain a stronger ($\ep$-independent) version
of the relative entropy-QAEP.
Similarly, we provide bounds on the one-shot $\ep$-error classical capacity of a quantum channel in terms of a smooth max-relative entropy variant of its Holevo capacity.
Using these bounds and the $\ep$-independent version
of the relative entropy-QAEP, we can recover both the Holevo-Schumacher-Westmoreland theorem about the optimal direct rate of a memoryless quantum channel with product state encoding, as well as its strong converse counterpart.
%
\end{abstract}
\maketitle

\section{Introduction}
Transmission of information through noisy channels is an essential requirement in various information-processing
tasks. A channel can be characterized by its capacity, which quantifies the maximum amount of information which
can be transmitted reliably per use of the channel. If the sender (Alice) encodes information at a rate less than the
capacity, then the receiver (Bob) can recover the information with a probability of error which vanishes asymptotically
in the number of uses of the channel. For rates above the capacity, the asymptotic probability of error is bounded
away from zero. Another quantity of interest characterizing a channel is its strong converse capacity, which is the rate
threshold above which information transmission fails with certainty, in the sense that the asymptotic probability of
error is equal to one.
Wolfowitz \cite{wolfo} proved that, for a memoryless classical channel, i.e., a classical channel for which there are no
correlations in the noise acting on successive inputs, the strong converse capacity is equal to the capacity. This is referred
to as the strong converse property (see e.g.~\cite{han}). The capacity of the channel hence provides a sharp threshold on
its information-carrying power. 

The classical capacity of memoryless quantum channels (with weak converse) was shown to be equal to the regularized Holevo capacity in 
\cite{Holevo,SW}. If codewords are restricted to product state inputs then regularization is not necessary, and the capacity is equal to 
the single-shot Holevo capacity of the channel.
Moreover, in this case the strong converse property holds, as
was
proved independently by Ogawa and Nagaoka \cite{ON}, and by Winter \cite{Winter}. 
It is known that the classical capacity with general inputs can be strictly larger than the product-state capacity \cite{Hastings}, and it is an open question whether the strong converse property still holds in this case. 
The only known result in this direction was obtained recently by 
K\"onig and Wehner \cite{KW}, who proved the
strong converse property 
for the unconstrained classical capacity
of a class of
quantum channels for which the Holevo capacity 
is additive. These include all unital qubit channels, the $d$-dimensional
depolarizing channel and the Werner-Holevo channel.

It is well known that the problems of sending classical information through a quantum 
channel and binary state discrimination (hypothesis testing) are closely related to each 
other; in particular, the direct part of the channel coding theorem \cite{Holevo,SW} can be obtained from the direct part of Stein's lemma
\cite{HN,ON3}. Moreover, the direct part of the quantum analogue of fixed-length source compression is an immediate consequence 
of the direct part of Stein's lemma.
The optimal asymptotic direct and strong converse rates coincide for binary state discrimination in the i.i.d.~case, and are 
equal to the relative entropy of the two states \cite{HP,ON2}.
Coding theorems are typically obtained in two steps:
\begin{enumerate}
\item
Establishing a trade-off relation between the rate and the error for finite $n$ (where $n$ denotes the number of channel uses, or the number of copies of the states in the above examples). These trade-off relations are given in terms of some entropic quantities. 
\item
Evaluating the asymptotics of these entropic quantities in the $n\to\infty$ limit to obtain the limiting optimal rate. 
\end{enumerate}

One standard way to do this is to use R\'enyi relative entropies or related quantities for the trade-off relations, and 
obtain the asymptotics by 
using additivity properties of these quantities
(see, e.g., \cite{Aud,Hayashi,KW,ON,ON2,Nagaoka,NSz}), or by expressing the asymptotic rate as a regularized entropic quantity.
Another approach, which has gained a lot of popularity recently, is to use smooth entropies and related quantities.
Smooth entropies for quantum information theory were introduced in \cite{RennerPhD}, 
and their theory further developed in   
\cite{Datta:2009cl,Datta:2009fy,AEP,TCR2,marco_thesis}. Smooth entropies interpolate between the operational and the entropic sides of the coding problems, and hence provide a different insight into these poblems compared to the R\'enyi entropy approach. Due to this interpolation property, the finite-size trade-off relations are typically more straightforward to obtain, and the asymptotic results follow by the application of robust, problem-independent techniques, like the so-called Quantum Asymptotic Equipartition Property (QAEP) \cite{AEP,marco_thesis}. 
This not only enables a unified treatment of many coding problems, but the techniques developed on the way
provide a new set of tools to attack such problems; see, for instance the recent result about the strong converse of the quantum capacity of degradable channels \cite{MWinter}.

In this paper, we show how the smooth entropy approach can be used to obtain the direct and the strong converse capacities for classical information transmission through memoryless quantum channels with product encoding.
The structure of the paper is as follows. In Section \ref{sec:preliminaries}, we give the necessary technical background on 
smooth relative entropies. In Section \ref{sec:hypotesting}, we derive two-sided bounds on the optimal type II error of quantum hypothesis testing in terms of the smoothed max-relative entropy of the two states representing the two hypotheses (Theorem \ref{prop:hypotesting bounds}). 
Using a suitable version of the QAEP (Corollary \ref{cor:smooth limit}), we can derive the strong converse rate for the asymptotic hypothesis testing problem with i.i.d.~hypotheses (Theorem \ref{thm:Stein converse}). On the other hand, 
when combined with a recent result on finite-size corrections in Stein's lemma \cite{AMV}, the bounds of Theorem \ref{prop:hypotesting bounds} yield a strengthening (an $\ep$-independent version) 
of Corollary \ref{cor:smooth limit} (Theorem \ref{thm:maxent expansion}). (A similar result appeared recently in \cite{TH}, after the submission of the first 
version of this paper.)
In Section \ref{sec:one-shot}, we consider a slightly more general channel model than usual quantum channels, and
give two-sided bounds on the $\ep$-error capacity for one single use of such a channel. These bounds are given in terms of a generalization of the Holevo capacity, where the
correlations are measured by the smooth max-relative entropy instead of the usual relative entropy. 
Using Theorem \ref{thm:maxent expansion}, we show
in Section \ref{sec:asymptotics} that these bounds are asymptotically tight in the sense that one can recover from them the asymptotic direct and strong converse capacities of a memoryless channel. In particular, we get the direct and strong converse capacities of a quantum channel with product-state encoding. 
We conclude in Section \ref{Sec_Discussion} by comparing our results to related results in the literature.

\section{{Preliminaries}}\label{sec:preliminaries}

For a Hilbert space $\cH$, let
${\cal B}(\cH)$ denote the algebra of linear operators acting on $\hil$,
let
$\B(\hil)_+$ denote the set of positive semi-definite operators on $\hil$, and let
${\cD}({\cH}) \subset {\cal B}(\cH)_+$ denote the set of density matrices (or states), i.e., 
positive semi-definite operators of unit trace. 
Unless otherwise stated, we assume all Hilbert spaces to be finite-dimensional.

For self-adjoint operators $A,B\in\B(\hil)$, let $\{A\ge B\}$ denote the spectral projection 
of $A-B$ corresponding to the interval $[0,+\infty)$; the spectral projections
$\{A>B\},\,\{A\le B\}$ and $\{A<B\}$ are defined in a similar way.
For a self-adjoint operator $A\in\B(\hil)$, we use the notations $A_+=A\{A>0\}$ and
$A_-=A\{A<0\}$ for its positive and negative parts, respectively. 
Note that for any self-adjoint $A,X\in\B(\hil)$ such that $0\le X\le I$, we have  
\begin{equation}\label{trace bound}
\Tr XA=\Tr XA_+-\Tr XA_-\le \Tr XA_+\le\Tr A_+. 
\end{equation}

The \ki{trace distance} between two operators $A$ and $B$ is given by
\begin{equation}\nonumber
\norm{A-B}_1 := \Tr|A-B|=\Tr \bz(A-B)_++(A-B)_-\jz.
\end{equation}
For $\rho,\sigma\in\B(\hil)_+$, their {\em fidelity} is
\begin{equation*}
F(\rho,\sigma):=\Tr\sqrt{\sqrt{\rho}\sigma\sqrt{\rho}}=\max_{\psi_\rho,\psi_\sigma}|\inner{\psi_\rho}{\psi_\sigma}|,
\end{equation*}
where the last expression is due to Uhlmann's theorem \cite{Uhlmann}, and the maximum is taken over all purifications $\psi_\rho,\psi_\sigma$ of $\rho$ and $\sigma$, respectively. It is easy to see that 
\begin{equation*}
\d(\rho,\sigma):=\min_{\psi_\rho,\psi_\sigma}\half\norm{\pr{\psi_\rho}-\pr{\psi_\sigma}}_1=\sqrt{\bz\Tr \rho+\Tr \sigma\jz^2/4-F(\rho,\sigma)^2},
\end{equation*}
and that $\d$ is a metric on $\B(\hil)_+$. The same arguments as in \cite{FvdG,Uhlmann} (see also \cite{nielsen}) yield that 
\begin{align*}
\frac{\d(\rho,\sigma)^2}{\Tr \rho+\Tr \sigma}\le\half\bz\Tr \rho+\Tr \sigma\jz-\sqrt{\bz\Tr \rho+\Tr \sigma\jz^2/4-\d(\rho,\sigma)^2} \le\half\norm{\rho-\sigma}_1\le \d(\rho,\sigma)
\end{align*}
(where the first expression should be replaced by $0$ if $\rho=\sigma=0$). For density operators $\rho,\sigma\in\D(\hil)$, the above expressions simplify to 
\begin{equation}\label{fuchs}
\d(\rho,\sigma)=\sqrt{1-F^2(\rho,\sigma)},\ds\ds\ds\text{and}\ds\ds\ds
\half \d(\rho,\sigma)^2\le  1-F(\rho,\sigma) \leq \frac{1}{2} \norm{\rho - \sigma}_1 \leq \d(\rho,\sigma).
\end{equation}
The distance $d_s(\rho,\sigma):=\sqrt{1-F^2(\rho,\sigma)}$ on density operators was introduced in \cite{GLN} under the name {\em sine distance}, and 
our definition provides a natural extension of it to the set of positive semidefinite operators. The sine distance
was extended to a metric on subnormalized states in a different way under the name {\em purified distance} in \cite{TCR2}.
To distinguish it from the purified distance while reflecting the fact that it is the minimal distance of purifications, we will use the terminology
{\em distance of optimal purifications}.


For $\rho,\sigma\in\B(\hil)_+$ and $\alpha\in(1,+\infty)$, the \ki{R\'enyi $\alpha$-relative entropy} of $\rho$ with respect to $\sigma$ is
 \begin{equation*}
 \rsr{\rho}{\sigma}{\alpha}:=
 \begin{cases}
 \frac{1}{\alpha-1}\log\Tr\rho^{\alpha}\sigma^{1-\alpha},&\supp\rho\subseteq\supp\sigma,\\
 +\infty,&\text{otherwise}.
 \end{cases}
 \end{equation*}
Here and henceforth logarithms are taken to base $2$. It is easily seen that $\alpha\mapsto\rsr{\rho}{\sigma}{\alpha}$ is monotone increasing for fixed $\rho$ and $\sigma$, and if $\rho$ is a density operator then $\lim_{\alpha\searrow 1}\rsr{\rho}{\sigma}{\alpha}=\sr{\rho}{\sigma}$, where $\sr{\rho}{\sigma}$ is the relative entropy, defined as
 \begin{equation}
 \sr{\rho}{\sigma}:=
 \begin{cases}
 \Tr\rho(\log\rho-\log\sigma),&\supp\rho\subseteq\supp\sigma,\\
 +\infty,&\text{otherwise}.
 \end{cases}
\label{qrel}
 \end{equation}
The \ki{von Neumann entropy} of a state $\rho$ is given by $S(\rho)= -\tr (\rho \log \rho)$.

\begin{lemma}\label{lem3}
Given a state $\rho_{AB}\in\D(\hil_{AB})$, let $\rho_{A(B)}=\tr_{B(A)}\rho_{AB}$. Then for any operator $\sigma_A\in\B(\hil_A)_+$,
\begin{equation}
\min_{\omega_B\in\D(\hil_B)} D(\rho_{AB}\|\sigma_A\otimes\omega_B)
=D(\rho_{AB}\|\sigma_A\otimes\rho_B).
\label{ab}
\end{equation}
\end{lemma}
\begin{proof}
Note that the left-hand side of \reff{ab} is equal to $+\infty$ if and only if 
the right-hand side is equal to $+\infty$. When both sides are finite, 
the assertion follows immediately from $\sr{\rho_{AB}}{\sigma_A\otimes\omega_B}-\sr{\rho_{AB}}{\sigma_A\otimes\rho_B}=
\sr{\rho_B}{\omega_B}\ge 0$ (see Lemma 6 in \cite{qcap} for more details).
\end{proof}

The notion of the (smoothed) min-entropy was introduced in \cite{RennerPhD}, 
which in our terminology would correspond to the conditional version of 
the (smoothed) max-relative entropy. The relative entropy version of the min-entropy has been introduced in 
\cite{Datta:2009fy} under the name of (smoothed) max-relative entropy, which is defined as follows. Note also that various symmetrised versions
of the max-relative entropy have been known in operator theory as the Hilbert projective metric \cite{Hilbert} and the Thompson distance \cite{Thompson}.

\begin{definition}\label{DEF_MAX_RELATIVE}
The max-relative entropy of two positive semi-definite operators $\rho$ and $\sigma$ is defined as
\begin{equation*}
D_{\max}(\rho\|\sigma):= {\inf\{\gamma:\,\rho\leq 2^\gamma\sigma\}}.
\end{equation*}
For any $0\leq\eps\leq 1$, the $\eps$-smooth max-relative entropy of {a state $\rho$ and a positive semi-definite operator $\sigma$} is defined as
\begin{equation*}
D^\eps_{\max}(\rho\|\sigma):=\min_{\bar{\rho}\in B_\eps(\rho)} D_{\rm{max}}(\bar{\rho}\|\sigma),
\end{equation*}
where
\begin{equation}\label{ball}
B_\eps(\rho):=\{\bar{\rho}\ge 0, \tr\bar{\rho}=1; \d(\bar\rho,\rho)\le\ep\}
\end{equation}
is the $\ep$-ball around $\rho$ with respect to $\d$.
\end{definition}

\begin{remark}
There are various, slightly different, ways to define the smoothing in the literature.
One common choice is to use the purified distance $d_p$ \cite{TCR2} instead of $\d$, and allow subnormalized states in the sense of replacing $B_\ep(\rho)$ in \eqref{ball} with $\tilde B_\eps(\rho)=\{\bar{\rho}\ge 0, \tr\bar{\rho}\le 1; d_p(\bar\rho,\rho)\le\ep\}$.
 We give some comments on our choice and its relation to that of \cite{TCR2} in Section
\ref{Sec_Discussion}.
\end{remark}
\medskip

Note that for $\ep=0$ we have $\srmaxs{\rho}{\sigma}{0}=\srmax{\rho}{\sigma}$, the 
function $\ep\mapsto\srmaxs{\rho}{\sigma}{\ep}$ is monotone decreasing,
and $\srmaxs{\rho}{\sigma}{\ep}=0$ if and only if $\ep\ge\d(\rho,\sigma)$.
A quantitative bound on the 
effect of smoothing is given by the following:

\begin{lemma}\label{lemma:smoothing-bound}
Let $\rho\in\D(\hil)$ and $\sigma\in\B(\hil)_+$ be such that $\rho\sigma\ne 0$, let $\lambda>0$, and let
$\Delta_+(\lambda) := \bigl(\rho- \lambda \sigma\bigr)_+$ be the positive part of the operator $(\rho- \lambda \sigma)$. Then
\begin{equation}\label{smoothing-bound}
D_{\max}^{\ep(\lambda)}(\rho\|\sigma) \le \log \frac{\lambda}{\sqrt{1-\ep(\lambda)^2}},\ds\ds
\text{where}\ds\ds \eps(\lambda): =\sqrt{\Tr \Delta_+(\lambda)\bz 2-\Tr \Delta_+(\lambda)\jz}.
\end{equation}

The function $\lambda\mapsto\Tr \Delta_+(\lambda)$ is convex on the whole real line.
If, moreover, $\supp\rho\le\supp\sigma$ then the function $\lambda\mapsto\Tr \Delta_+(\lambda)$ is strictly decreasing 
and continuous
on $\left[0,2^{D_{\max}(\rho\|\sigma)}\right]$ with range $[0,1]$, 
and the function 
\begin{equation*}
\lambda\mapsto \ep(\lambda)=\sqrt{\Tr \Delta_+(\lambda)\bz 2-\Tr \Delta_+(\lambda)\jz}
\end{equation*}
is strictly decreasing and continuous on $\left[0,2^{D_{\max}(\rho\|\sigma)}\right]$ with range $[0,1]$.
\end{lemma}
\begin{proof}
By definition, $\rho\le\lambda\sigma+\Delta_+(\lambda)$. Using Lemma C.5 in \cite{BP}
(see also Lemma 5 in \cite{Datta:2009cl}), we obtain the existence of a state $\widetilde\rho$ such that 
$\widetilde\rho\le(1-\Tr\Delta_+(\lambda))^{-1}\lambda\sigma$ and
$F(\rho,\widetilde\rho)\ge 1-\Tr\Delta_+(\lambda)$. By the former, we have
$\srmax{\widetilde\rho}{\sigma}\le \log\lambda(1-\Tr\Delta_+(\lambda))^{-1}$, and by the latter,
$\sqrt{1-F^2(\rho,\widetilde\rho)}\le \ep(\lambda)$. Thus,
$\widetilde\rho\in B_{\ep(\lambda)}$, and hence
$\srmaxs{\rho}{\sigma}{\ep(\lambda)}\le\srmax{\widetilde\rho}{\sigma}$, from which 
\eqref{smoothing-bound} follows.

Let $\lambda_0,\lambda_1>0$ and let $\lambda_p:=(1-p)\lambda_0+p\lambda_1$ for every 
$p\in[0,1]$. Then
\begin{align*}
\Tr\left[\Delta_+(p\lambda_0 +(1-p)\lambda_1)\right]&=
\Tr\left[\{\rho-\lambda_p\sigma>0\}(\rho-\lambda_p\sigma)\right]\\
&=\Tr\left[\{\rho-\lambda_p\sigma>0\}\bigl[(1-p)(\rho-\lambda_0\sigma)
+p(\rho-\lambda_1\sigma)\bigr]\right]\\
&=(1-p)\Tr\left[\{\rho-\lambda_p\sigma>0\}(\rho-\lambda_0\sigma)
+p\Tr\{\rho-\lambda_p\sigma>0\}(\rho-\lambda_1\sigma)\right]\\
&\le
(1-p)\Tr\left(\Delta_+(\lambda_0)\right)+p\Tr\left(\Delta_+(\lambda_1)\right),
\end{align*}
where the last inequality follows from \eqref{trace bound}. This proves the 
assertion on the convexity. 

Assume now that $\supp\rho\le\supp\sigma$. Note that $\Tr\Delta_+(\lambda)=0$ if and only if
$ \lambda\ge 2^{D_{\max}({\rho}\|{\sigma})}$.
If $\supp\rho\le\supp\sigma$ then $D_{\max}({\rho}\|{\sigma})<+\infty$, and, since
$\Tr\Delta_+(\lambda)>0$ when  $\lambda< 2^{D_{\max}({\rho}\|{\sigma})}$,
convexity yields that
$\lambda\mapsto\Tr\Delta_+(\lambda)$ is strictly decreasing on 
$(-\infty, 2^{D_{\max}({\rho}\|{\sigma})}]$. 
In particular, it is strictly decreasing on $\left[0,2^{D_{\max}(\rho\|\sigma)}\right]$, with range $[0,1]$, and convexity implies that it is also continuous. Since 
$x\mapsto \sqrt{x(2-x)}$ is strictly increasing and continuous on $[0,1]$ with range $[0,1]$, the statement follows.
\end{proof}

\begin{remark}
Note that $\Tr\Delta_+(\lambda)=1$ for $\lambda=0$, and 
the convexity of $\lambda\mapsto\Tr\Delta_+(\lambda)$ yields that 
$\lambda\mapsto\frac{\Tr\Delta_+(\lambda)-1}{\lambda}=\frac{-\sqrt{1-\ep(\lambda)^2}}{\lambda}$ is monotone increasing, and hence so is the upper bound in \eqref{smoothing-bound}.
\end{remark}

\begin{remark}
A bound similar to our inequality \eqref{smoothing-bound} appeared in Lemma 6.1 of \cite{marco_thesis}. The difference between the two 
is the extra factor $\sqrt{1-\ep(\lambda)^2}$ in our bound, which is due to our different choice of smoothing.
\end{remark}

\bigskip

Let $\rho_x,\sigma_x\in\B(\hil)_+$ for every $x\in\X$, where $\X$ is a finite set, let
$\{p_x\}_{x\in\X}$ be a probability distribution on $\X$ and let $\{\pr{x}\}_{x\in\X}$ be a set of orthonormal rank-$1$ projections on some Hilbert space $\kil$. It follows immediately from Definition \ref{DEF_MAX_RELATIVE} that
\begin{align}\label{dmax classical-quantum}
D_{\max}\bz\sum_{x\in\X} p_x\rho_x \Big\| \sum_{x\in\X} p_x\sigma_x\jz\nonumber
&\le
\max_{x:\,p_x>0} \srmax{\rho_x}{\sigma_x}\nonumber\\
&=
D_{\max}\bz\sum_{x\in\X} p_x\pr{x}\otimes\rho_x \Big\| \sum_{x\in\X} p_x\pr{x}\otimes\sigma_x\jz.
\end{align}
The first inequality says that the max-relative entropy is jointly
quasi-convex in its arguments (see also Lemma 9 in \cite{Datta:2009fy}).
This in turn implies joint quasi-convexity of the $\ep$-smooth max-relative entropy:
\begin{lemma}\label{lemma:quasi}
For any $0\leq\eps\leq 1$,
\begin{equation}\label{quasi_eq}
D^\eps_{\max}\bz\sum_i \gamma_i \rho_i\Big\|\sum_i \gamma_i \sigma_i\jz
\le \max_i  D^\eps_{\max}(\rho_i\|\sigma_i),
\end{equation}
where for each $i$, $\gamma_i >0$, $\rho_i, \sigma_i$ are states, and $\sum_i \gamma_i=1$.
\end{lemma}
\begin{proof}
{For every $i$, let $\nu_i\in B_\eps(\rho_i)$ such that
\begin{equation}\label{quasi_eq3.1}
D_{\max}^\eps(\rho_i\|\sigma_i)=D_{\max}(\nu_i\|\sigma_i).
\end{equation}
Due to the joint concavity of the fidelity \cite{nielsen}, we have
\begin{align*}
F\bz\sum_i \gamma_i \nu_i,\sum_i \gamma_i \sigma_i\jz&\ge
\sum_i \gamma_i F(\nu_i,\sigma_i)\ge
\sum_i \gamma_i \sqrt{1-\ep^2}=\sqrt{1-\ep^2},
\end{align*}
i.e., $\sum_i \gamma_i \nu_i\in \cB_\ep(\sum_i \gamma_i \sigma_i)$. Thus,
\begin{align*}
D_{\max}^\eps\bz\sum_i \gamma_i \rho_i\Big\|\sum_i \gamma_i \sigma_i\jz&\le
D_{\max}\bz\sum_i \gamma_i \nu_i\Big\| \sum_i \gamma_i \sigma_i\jz\\
&\le
\max_i  D_{\max}(\nu_i\|\sigma_i)
=
\max_i D^{\ep}_{\max}(\rho_i\|\sigma_i),
\end{align*}
where in the second inequality we used the quasi-convexity of the max-relative entropy.}
\end{proof}


Lemma \ref{lemma:BP} and Lemma \ref{key2} given below relate the smoothed max-relative entropy to the quantum relative entropy and the $\alpha$-relative entropies, respectively:

\begin{lemma}\label{lemma:BP}
Let $\rho_{AB}\in\D(\hil_{AB})$ be a bipartite state and let $\sigma_A\in\D(\hil_A)$ be such that $\supp\rho_A\subseteq\supp\sigma_A$. Then
\begin{equation*}
\lim_{\eps\to 0}\limsup_{n\to\infty}\frac{1}{n}
\min_{\omega_n\in\D(\hil_B^{\otimes n})}D^{{\eps}}_{\rm{max}}(\rho_{AB}^{\otimes n}\|\sigma_A^{\otimes n} \otimes \omega_n)
= D(\rho_{AB}\|\sigma_A \otimes \rho_B).
\end{equation*}
\end{lemma}
\begin{proof}
The assertion follows immediately from Proposition II.1 in \cite{BP} by choosing $\M_n=\sigma_A^{\otimes n}\otimes\D(\hil_B^{\otimes n})$ for every $n\in\bN$, and by taking into account Lemma \ref{lem3}. Note that in \cite{BP}, it was assumed that $\M_1$ contains a state with full rank, but it is obviously enough to assume that 
$\M_1$ contains a state with support larger than or equal to that of $\rho_{AB}$. 
\end{proof}

By choosing system $B$ to be one-dimensional in the above lemma, we obtain the following:
 
 \begin{corollary}\label{cor:smooth limit}
 For states $\rho$ and $\sigma$ such that $\supp \,\rho \subseteq \supp\, \sigma$,
 \be\label{asympmax0}
  \lim_{\eps\to 0}\limsup_{n\to\infty}\frac{1}{n}
 D^{{\eps}}_{\rm{max}}(\rho^{\otimes n}\|\sigma^{\otimes n})
 = D(\rho\|\sigma).
 \ee
 \end{corollary}
A related result was obtained in Theorem 2 in \cite{Datta:2009fy}, where it was shown that the left-hand side of \eqref{asympmax0} is equal to the sup spectral divergence rate.
A conditional entropy version of the above Theorem was proved in \cite{AEP}, under the name of 
\textit{fully quantum asymptotic equipartition property}; see also \cite{marco_thesis}.

\begin{lemma}\label{key2}
Let 
$\rho\in\D(\hil)$, $\sigma\in\B(\hil)_+$,
$\eps\in(0,1)$ and $\alpha \in (1,2]$. Then
\begin{equation}\label{smooth-Renyi inequality}
D_{\max}^\eps(\rho\|\sigma) \le
D_{\alpha}(\rho\|\sigma)
+\frac{1}{\alpha-1} \log \frac{2}{\ep^2}
-\log\sqrt{1-\ep^2}
\end{equation}
\end{lemma}
\begin{proof}
The proof is exactly analogous to that of Theorem 7 in \cite{AEP}. Let $\lambda>0$ be such that
$\sqrt{1-\ep^2}=1-\Tr(\rho-\lambda\sigma)_+$ (cf.~Lemma \ref{lemma:smoothing-bound}). Following the proof of Theorem 7 in \cite{AEP}, we obtain
\begin{equation*}
1-\sqrt{1-\ep^2}=\Tr(\rho-\lambda\sigma)_+
\le\lambda^{1-\alpha}\Tr\rho^\alpha\sigma^{1-\alpha},
\end{equation*}
and using Lemma \ref{lemma:smoothing-bound} we get
\begin{align*}
D_{\max}^\eps(\rho\|\sigma)& \le\log\frac{\lambda}{\sqrt{1-\ep^2}}\le \rsr{\rho}{\sigma}{\alpha}+ \frac{1}{\alpha-1} \log \bz 1-\sqrt{1-\ep^2}\jz\inv-\log\sqrt{1-\ep^2}.
\end{align*}
Using that $\bz 1-\sqrt{1-\ep^2}\jz\inv\le 2/\ep^2$, we finally get \eqref{smooth-Renyi inequality}.
\end{proof}

\bigskip

%
\section{Quantum Hypothesis Testing}\label{sec:hypotesting}

Consider the quantum hypothesis testing problem with the null hypothesis $H_0:\rho$ versus the alternative hypothesis $H_1:\sigma$,
where $\rho$ and $\sigma$ are density operators on some finite-dimensional Hilbert space $\hil$. We can decide which hypothesis is true based on the POVM $\{\Pi,I-\Pi\}$, where $0\leq \Pi \leq I$. For a test $\Pi$, the error 
probability of the first kind (or type I error) and the second kind (or type II error)
are defined as
\begin{eqnarray}
\alpha(\Pi)&:=&\tr [(I-\Pi)\rho], \\
\beta(\Pi)&:=&\tr[\Pi\sigma],
\end{eqnarray}
respectively,
where $\alpha(\Pi)$ is the probability of accepting $\sigma$ when $\rho$ is true while $\beta(\Pi)$ is the probability of accepting $\rho$ when $\sigma$ is true. 
Obviously, there is a trade-off between the two error probabilities, and there are various ways to jointly optimize them.
In the asymmetric setting of Stein's lemma \cite{CT,HP,ON2}, the error probability of the second kind is optimized under the constraint that the 
error probability of the first kind stays below a threshold $\ep\in(0,1)$; the optimal error of the second kind is then given by
\begin{equation*}
\beta_{\ep}(\rho\|\sigma):=\min\{\beta(\Pi)\,:\,\alpha(\Pi)\le \ep\},
\end{equation*}
where the minimization is over all POVMs $\{\Pi, I-\Pi\}$. In general, there is no closed formula known 
for $\beta_\ep(\rho\|\sigma)$ or for the optimal POVM attaining it. However, we can give the following bounds in 
terms of the smoothed max-relative entropy of $\rho$ and $\sigma$:

\begin{theorem}\label{prop:hypotesting bounds}
Assume that $\supp\rho\subseteq\supp\sigma$.
For any $0<\ep'<\eps<1$, 
\begin{equation*}
D^{g(\ep)}_{\rm{max}}(\rho\|\sigma)
\le
-\log\beta_{1-\ep}(\rho\|\sigma)
\le
D^{\ep'}_{\max}(\rho\|\sigma)+\log \frac{1}{\ep-\ep'},
\end{equation*}
where $g(\ep):=\sqrt{\eps(2-\ep)}$.
\end{theorem}
\begin{proof}
\textit{(Upper bound)}\ds
Let $\ep\in(0,1)$ be fixed.
The assertion will follow if we can show that for any $0 \le \Pi \le I$ such that
\begin{equation}\label{UPP2}
\log\beta(\Pi)< -D^{\ep'}_{\rm{max}}(\rho\|\sigma)-\log \frac{1}{\ep-\ep'}
\end{equation}
we have 
\begin{equation*}
\alpha(\Pi) > 1 - \eps.
\end{equation*}

Thus, let $\Pi$ be such that \eqref{UPP2} holds.
By the definition of $D^{\ep'}_{\rm{max}}(\rho\|\sigma)$, there exists a state $\bar{\rho}\in B_{\ep'}(\rho)$ for which
\begin{equation}\label{UPP1}
\bar{\rho}\leq 2^{D^{\ep'}_{\rm{max}}(\rho\|\sigma)}\sigma,
\end{equation}
and therefore
\begin{eqnarray}
\tr \Pi\bar{\rho} &\leq& 2^{D^{\ep'}_{\rm{max}}(\rho\|\sigma)} 
\tr (\Pi\sigma) \nonumber\\
&=&2^{D^{\ep'}_{\rm{max}}(\rho\|\sigma)} \beta(\Pi) \nonumber\\
&<& 2^{D^{\ep'}_{\rm{max}}(\rho\|\sigma)}2^{-D^{\ep'}_{\rm{max}}(\rho\|\sigma)+\log (\ep-\ep')}\nonumber\\
&=&\ep-\ep'. \label{UPP3}
\end{eqnarray}
The first inequality follows from (\ref{UPP1}), and the second inequality follows from (\ref{UPP2}).
Hence,
\begin{equation*}
1-\alpha(\Pi)=\tr (\Pi\rho)= \tr (\Pi\bar{\rho}) + \tr (\Pi (\rho-\bar{\rho}))
< \ep-\ep' + \|\rho-\bar{\rho}\|_1/2
\leq \eps, 
\end{equation*}
where the first inequality follows from (\ref{UPP3}) and the second inequality holds because $\bar{\rho}\in B_{\ep'}(\rho)$. 

\textit{(Lower bound)}\ds
By Lemma~\ref{lemma:smoothing-bound}, there exists a $\lambda>0$ such that 
$\Tr(\rho-\lambda\sigma)_+=\ep$.
For this $\lambda$, let $\Pi:=\{\rho\geq \lambda\sigma\}$. Then 
\begin{equation*}
\tr\Pi\rho\geq \tr\Pi(\rho-\lambda\sigma)=\tr (\rho-\lambda\sigma)_+=\eps,
\end{equation*}
or equivalently, $\alpha(\Pi)\leq 1-\eps$, and hence
\begin{equation*}
-\log\beta_{1-\ep}(\rho\|\sigma)\ge -\log\beta(\Pi).
\end{equation*}
On the other hand, 
$\ep= \Tr(\rho-\lambda\sigma)_+=\Tr\Pi(\rho-\lambda\sigma)\le 1-\lambda\Tr\Pi\sigma$
yields
\begin{eqnarray*}
\beta(\Pi)&=&\tr \Pi\sigma \leq \frac{1-\ep}{\lambda},
\end{eqnarray*}
and hence,
\begin{equation*}
-\log\beta(\Pi) \geq \log \lambda-\log(1-\ep) \geq D^{g(\ep)}_{\rm{max}}(\rho\|\sigma)+\log\sqrt{1-g(\ep)^2}-\log(1-\eps)
=
D^{g(\ep)}_{\rm{max}}(\rho\|\sigma),
\end{equation*}
where we have used Lemma~\ref{lemma:smoothing-bound}.
\end{proof}

The above bounds, combined with Corollary \ref{cor:smooth limit}, can be used to derive 
the strong converse theorem for hypothesis testing:

\begin{definition}
The asymptotic strong converse rate $R_{sc}$ of the quantum hypothesis testing problem for the null hypothesis $H_0:\rho$ versus the alternative hypothesis $H_1:\sigma$ is defined 
to be the smallest number $R$ such that if 
\begin{equation*}
\limsup_{n\to\infty}\frac{1}{n}\log\Tr\Pi_n\sigma^{\otimes n}\le -R
\end{equation*}
for some sequence of tests $\{\Pi_n\}_{n\in\bN}$ then 
{{\begin{equation*}
\lim_{n\to\infty}\Tr(I_n-\Pi_n)\rho^{\otimes n}=1.
\end{equation*}}}
\end{definition}

\begin{theorem}[\cite{ON2}]\label{thm:Stein converse}
The asymptotic strong converse rate 
$R_{sc}$ of the quantum hypothesis testing problem for the null hypothesis $H_0:\rho$ versus the alternative hypothesis $H_1:\sigma$
is given by
\begin{equation}\label{Stein sc}
R_{sc} = \sr{\rho}{\sigma},
\end{equation}
where $\sr{\rho}{\sigma}$ is the quantum relative entropy \eqref{qrel}.
\end{theorem}
\begin{proof}
It is easy to see that 
\begin{equation*}
R_{sc}=\lim_{\eps\to 0}\limsup_{n\to \infty}\frac{1}{n}\log \beta_{1-\eps}\bz\rho^{\otimes n}\|\sigma^{\otimes n}\jz.
\end{equation*} 
The assertion then follows from Theorem \ref{prop:hypotesting bounds} and Corollary \ref{cor:smooth limit}.
\end{proof}

In Theorem \ref{prop:hypotesting bounds} we have derived bounds on the optimal type II error in terms of the smoothed max-relative entropy, and used 
Corollary \ref{cor:smooth limit} to obtain the strong converse rate for Stein's lemma. Proceeding the other way around, we can use the bounds of Theorem \ref{prop:hypotesting bounds}
together with a recent result from \cite{AMV},
to obtain a significantly stronger version of 
Corollary \ref{cor:smooth limit}.
Indeed, 
the bounds in Theorem \ref{prop:hypotesting bounds} can be rewritten as
\begin{equation}\label{hypotesting bounds rewritten}
-\log\beta_{1-\ep'}(\rho\|\sigma)-\log\frac{1}{\ep'-\ep}
\le
D^{\ep}_{\max}(\rho\|\sigma)\le
-\log\beta_{\sqrt{1-\ep^2}}(\rho\|\sigma)
\end{equation}
for every $0<\ep<\ep'<1$.
Theorem 3.3 in \cite{AMV} says that for every $\ep\in(0,1)$ and $n\in\bN$,
\begin{equation*}
\sr{\rho}{\sigma}-\frac{f_1(\ep)}{\sqrt{n}}
\le
-\frac{1}{n}\log\beta_{1-\ep}\bz\rho^{\otimes n}\|\sigma^{\otimes n}\jz
\le
\sr{\rho}{\sigma}+\frac{f_2(\ep)}{\sqrt{n}},
\end{equation*}
where $f_1(\ep),f_2(\ep)>0$ are defined as $f_1(\ep):=4\sqrt{2}\log(1-\ep)\inv\log\eta,\,f_2(\ep):=4\sqrt{2}\log\ep\inv\log\eta$ and
$\eta:=1+\Tr\rho^{3/2}\sigma^{-1/2}+\Tr\rho^{1/2}\sigma^{1/2}$. Comparing it with 
\eqref{hypotesting bounds rewritten},
we obtain the following:

\begin{theorem}\label{thm:maxent expansion}
For every $\rho,\sigma\in\D(\hil)$ such that $\supp\rho\le\supp\sigma$,
every $0<\ep<\ep'<1$, and every $n\in\bN$, we have 
\begin{align}
\frac{1}{n}D_{\max}^\ep\bz\rho^{\otimes n}\|\sigma^{\otimes n}\jz
&\le
\sr{\rho}{\sigma}+\frac{1}{\sqrt{n}}4\sqrt{2}(\log\eta)\log(1-\sqrt{1-\ep^2})\inv,
\label{relentr bounds2}\\
\frac{1}{n}D_{\max}^\ep\bz\rho^{\otimes n}\|\sigma^{\otimes n}\jz
&\ge
\sr{\rho}{\sigma}-\frac{1}{\sqrt{n}}4\sqrt{2}(\log\eta)\log(1-\ep')\inv
-\frac{1}{n}\log\frac{1}{\ep'-\ep}.\label{relentr bounds3}
\end{align}
In particular,
\begin{equation}\label{strong AEP}
\lim_{n\to\infty}\frac{1}{n}D_{\max}^\ep\bz\rho^{\otimes n}\|\sigma^{\otimes n}\jz
=
\sr{\rho}{\sigma}\ds\ds\ds
\text{for every}\ds\ep\in(0,1).
\end{equation}
\end{theorem}

\begin{remark}
An analogy of the upper bound \eqref{relentr bounds2} has been obtained before in \cite{AEP} for conditional entropies, and it was extended 
to relative entropies in \cite{marco_thesis}, where the upper bound 
\begin{align}\label{Marco bound}
\frac{1}{n}\tilde D_{\max}^\ep\bz\rho^{\otimes n}\|\sigma^{\otimes n}\jz
&\le
\sr{\rho}{\sigma}+\frac{1}{\sqrt{n}}4(\log\eta)\sqrt{\log(1-\sqrt{1-\ep^2})\inv},
\end{align}
was obtained for all $n\ge \frac{8}{5}\log(1-\sqrt{1-\ep^2})\inv$. Here, $\tilde D_{\max}^\ep$ is the smoothed max-relative entropy 
according to the smoothing convention of \cite{TCR2}; see Section \ref{Sec_Discussion} for its definition and its relation to our
$D_{\max}^\ep$. The difference between the two definitions yields a correction of order $1/n$, which is negligible compared to the 
$1/\sqrt{n}$ term.
Note that the $\log(1-\sqrt{1-\ep^2})\inv$ is under the square root in \eqref{Marco bound}, which is better than in
\eqref{relentr bounds2} when $\ep<\sqrt{3}/2$ and worse for $\ep>\sqrt{3}/2$. On the other hand, \eqref{relentr bounds2} holds for every $n\in\bN$, while
\eqref{Marco bound} only holds for large enough $n$, depending on $\ep$.
 
The exact second order asymptotics of $-\log\beta_{\ep}(\rho^{\otimes n}\|\sigma^{\otimes n})$, i.e, the limit
\begin{equation*}
\lim_{n\to+\infty}\sqrt{n}\bz -\frac{1}{n}\log\beta_{\ep}(\rho^{\otimes n}\|\sigma^{\otimes n})-D(\rho\|\sigma)\jz
\end{equation*}
has been evaluated very recently in \cite{Li}, and independently in \cite{TH}.
For large $n$, this yields sharper bounds than 
the ones in Theorem \ref{thm:maxent expansion}.
The advantage of the bounds in Theorem \ref{thm:maxent expansion} is, however, that they hold for every $n\in\bN$, and hence they provide easily computable bounds for any finite value of $n$.

The limit relation \eqref{strong AEP} has also been obtained in the recent paper \cite{TH}.
\end{remark}

\section{One-shot capacity for transmission of classical information}
\label{sec:one-shot}

A quantum channel is usually defined as a CPTP (completely positive and trace-preserving) linear map from $\D(\hil_A)$ to $\D(\hil_B)$, where $\hil_A$ and $\hil_B$ are 
(finite-dimensional) Hilbert spaces. Here we consider a more general channel model, where by a \ki{channel} $W$
we mean a map $W:\,\X\to\D(\hil_B)$, where $\hil_B$ is a finite-dimensional Hilbert space, and $\X$ is an arbitrary set, with no particular assumption on its cardinality or any mathematical structure. Obviously, usual quantum channels form a special subclass of this channel model, where the input set $\X$ is chosen to be the state space of some finite-dimensional Hilbert space, and $W$ is assumed to be linear and CPTP.
The channel $W:\,\X\to\D(\hil_B)$ is {\em{classical}} if its image ${\hbox{ran }}W:=\{W(x)\}_{x\in\cX}$  is a commutative subset of ${\cal B}({\cal H}_B)$.

Suppose that Alice (the sender) wants to communicate with Bob (the receiver) using the channel $W$. To do this, they agree on a finite set of possible messages, labelled by natural numbers from $1$ to $M$. To send the message labelled by $m\in\{1,\ldots,M\}$, Alice has to encode her message into an input signal of the channel, $\vfi(m)\in\X$, and send it through the channel $W$, resulting in the quantum state $W(\vfi(m))$ at Bob's side. Bob then performs a POVM (positive operator-valued measure) $\Pi:=\{\Pi_i\}_{i=1}^{M}$, and if the outcome corresponding to 
$\Pi_k$ happens, he concludes that the message with label $k$ was sent. The probability of this event is $\Tr \left(W(\vfi(m))\Pi_k\right)$. 
A triple $(M,\vfi,\Pi)$, as above is called a \ki{code}. More precisely, a code $\C$ is a
triple $\C=(M,\vfi,\Pi)$, where
\begin{itemize}
\item
$M\in\bN$ is the number of possible messages;
\item 
$\varphi:\,\{1,2,\cdots,M\}\to\X$ is Alice's encoding of possible messages into input signals of the channel;
\item
$\Pi:= \{\Pi_m\}_{m=1}^M$ (with $\Pi_m \ge 0$ $\forall\, m =1,2, \ldots M$, and $\sum_{m=1}^M \Pi_m=I$) is a POVM on $\hil_B$, performed by Bob to identify the message (decoding).
\end{itemize}
%

\noindent The \ki{average error probability} $p_e(\cC,W)$ of a code $\cC=(M,\varphi,\Pi)$ is defined as
\begin{equation}\label{perr}
p_e(\cC,W):=\frac{1}{M}\sum_{i=1}^M \left[1- \Tr \left(W(\varphi(i))\Pi_i\right)\right].
\end{equation}
\begin{definition}\label{one-shot-capacity}
For a given $\eps > 0$, the {\textbf{one-shot $\eps$-error capacity}}, $C_\eps^{(1)}(W)$, of a channel
$W$ is defined as follows:
\begin{equation}\label{cap1}
C_\eps^{(1)}(W):=\sup \{\log M: \, \exists \,\cC:=(M,\varphi,\Pi)\ {\rm{s.t.}}\ p_e(\cC, W)\leq \eps\}.
\end{equation}
\end{definition}
Note that it denotes the maximum number of bits that can be transmitted through a single
use of the channel with average error probability of at most $\eps$.

Our aim is to give bounds on the above defined operational capacities in 
terms of entropic quantities. To this end, we will need the notions of the $\alpha$-capacities and $\ep$-max capacities of a channel, which we define below.

For a set $\X$, let $\P_f(\X)$ denote the set of finitely supported probability distributions on $\X$. Note that if $\X=\S(\hil_A)$ for some Hilbert space $\hil_A$ then specifying a $p\in\P_f(\S(\hil_A))$ is equivalent to specifying an ensemble of states 
$\{\rho_k,p_k\}_{k=1}^r$, where $\rho_k\in\S(\hil_A),\,p_k\ge 0,\,k=1,\ldots r$, and $p_1+\ldots+p_r=1$. For every set $\X$, let $\hil_{\X}$ be a Hilbert space with 
$\dim\hil_{\X}=|\X|$, and let $\{\ket{x}\}_{x\in\X}$, be an orthonormal basis in $\hil_{\X}$.
For any divergence measure $\M$, we define the corresponding capacity $\chi^*_{\M}(W)$ of 
a channel $W:\,\X\to\S(\hil_B)$ as 
\begin{equation}\label{cap def}
\chi^*_{\M}(W):=\sup_{p\in\P_f(\X)}\chi_{\M}(W,p),
\end{equation}
\begin{equation}\label{Holevo quantity}
\hbox{with } \quad \quad  \quad \quad \chi_{\M}(W,p):=\inf_{\sigma_B\in\S(\hil_B)}\M\bz\rho_{\X B}(p)\|\rho_{\X}(p)\otimes\sigma_B\jz,
\end{equation}
\begin{equation}\label{rho XB}
\hbox{where } \quad \quad \quad \quad  \rho_{\X B}(p):=\sum_{x\in\X}p(x)\pr{x}\otimes W(x),\ds\ds\ds p\in\P_f(\X),
\end{equation}
and $\rho_{\X}(p):=\Tr_B\rho_{\X B}(p)=\sum_{x\in\X}p(x)\pr{x}$.
Note that $\chi_{\M}(W,p)$ measures the amount of correlation in the classical-quantum state
$\rho_{\X B}(p)$, with respect to the divergence measure $\M$. 

In particular, the $\alpha$-capacities \cite{Csiszar,KW,MH} and the $\ep$-max capacities of a channel 
are defined by choosing $\M=D_\alpha$ and $\M=D_{\max}^{\ep}$ in \eqref{cap def}, respectively.
We use the short-hand notations $\chi_\alpha(W,p),\,\chi^*_\alpha(W)$, $\chi_{\max,\ep}(W,p)$ and $\chi^*_{\max,\ep}(W)$ for the corresponding quantities.
A quantity related to our $\chi_{\max,\ep}(W,p)$ appeared in \cite{Berta}, under the name \ki{smooth max-information}.
In the case of $\M=D_{\alpha}$, there is an explicit expression for the infimum in \eqref{Holevo quantity}, and for the optimal 
$\sigma_B$ achieving it; see, e.g., \cite{Csiszar,KW,Sibson}.

Lemma \ref{key2} yields the following inequality between the $\ep$-max capacity and the $\alpha$-capacities:
\begin{lemma}\label{lemma:ep-max alpha capacity bound}
For any channel $W$, any $\ep\in(0,1)$ and any $\alpha\in(1,2]$, we have 
\begin{equation*}
\chi_{\max,\ep}^*(W)\le\chi_{\alpha}^*(W)+\frac{1}{\alpha - 1} \log \frac{2}{\eps^2}-\log\sqrt{1-\ep^2}.
\end{equation*}
\end{lemma}
\bigskip

In the limit $\alpha\to 1$, the $\alpha$-capacities yield the Holevo capacity $\chi^*(W)$ \cite{MH,ON}:
\begin{equation}\label{alpha cap limit}
\lim_{\alpha\to 1}\chi^*_\alpha(W)=\chi^*(W):=\chi^*_D(W)=
\sup_{p\in\M_f(\X)}\sr{\rho_{\X B}(p)}{\rho_{\X}(p)\otimes\rho_B(p)},
\end{equation}
where $D$ stands for the relative entropy \eqref{qrel}.
\smallskip

The $\ep$-max capacity is quasi-convex as a function of the channel, as
is stated in the following lemma.
\begin{lemma}\label{lemma:max cap quasi}
Let $W_i:\,\X\to\S(\hil_B)$ be channels for $i=1,\ldots,r$, and let $\{\gamma_i\}_{i=1}^r$ be a probability distribution. For every $\ep\in[0,1]$,
%
\begin{equation}\label{max cap quasi2}
\chi^*_{\max,\ep}\bz\sum_i \gamma_i  W_i\jz
\le
\max_i \chi^*_{\max,\ep}\bz W_i\jz.
\end{equation}
\end{lemma}
\begin{proof}
Let $p\in\P_f(\X)$, $\rho_{\X B}^i:=\sum_{x\in\X}p(x)\pr{x}\otimes W_i(x)$ and
$\rho_{\X B}:=\sum_i\gamma_i\rho_{\X B}^i$. 
Note that $\rho_{\X}=\rho_{\X}^i=\sum_x p(x)\pr{x}$ for every $i$. 
For every $i$, let $\sigma_i\in\S(\hil_B)$ be such that $\chi_{\max,\ep}(W_i,p)=D_{\max}^{\ep}(\rho_{\X B}^i\|\rho_{\X}^i\otimes\sigma_i)$. Then
\bea
\chi_{\max,\ep}\bz\sum_i\gamma_i W_i,p\jz &\le&
D_{\max}^{\ep}\bz\rho_{\X B}\Big\|\rho_{\X}\otimes\sum_i\gamma_i\sigma_i\jz\nonumber\\
&=&
D_{\max}^{\ep}\bz\sum_i\gamma_i\rho_{\X B}^i\Big\|\rho_{\X}\otimes\sum_i\gamma_i\sigma_i\jz\nonumber\\
&\le&
\max_i D_{\max}^{\ep}\bz\rho_{\X B}^i\|\rho_{\X}\otimes\sigma_i\jz\nonumber\\
&=&
\max_i D_{\max}^{\ep}\bz\rho_{\X B}^i\|\rho_{\X}^i\otimes\sigma_i\jz\nonumber\\
&=&
\max_i\chi_{\max,\ep}\bz W_i,p\jz,\label{r1}
\eea
where the first inequality is due to the definition \eqref{cap def} and the second is due to Lemma \ref{lemma:quasi}. The inequality \eqref{max cap quasi2} follows immediately from \reff{r1}.
\end{proof}

After this preparation, we are ready to give the main result of the paper:

\begin{theorem}\label{thm:one-shot bounds}
For any $0<\ep'<\ep<\ep''<1$, the one-shot $\ep$-error
capacity of a channel $W$ satisfies the following bounds:
\begin{equation}\label{keybounds}
\chi^{*}_{\max,\sqrt{1-(\ep')^2}}(W)+
\log\frac{\ep'(\ep-\ep')^2}{8\ep}
\le 
C^{(1)}_{\eps}(W)
\le
\chi^*_{\max,1-\ep''}(W)-\log(\ep''-\ep).
\end{equation}
\end{theorem}


Before proving Theorem \ref{thm:one-shot bounds}, we give the following corollaries:

\begin{corollary}\label{cor:alpha upper bound}
In the setting of Theorem \ref{thm:one-shot bounds}, we have 
\begin{equation}\label{upper bound with alpha cap}
C^{(1)}_{\eps}(W)\le\chi^*_{\alpha}(W)+\frac{1}{\alpha - 1} \log \frac{2}{(1-\ep'')^2}
-\log(\ep''-\ep).
\end{equation}
\end{corollary}
\begin{proof}
Immediate from the second inequality in \eqref{keybounds} and Lemma \ref{lemma:ep-max alpha capacity bound}.
\end{proof}

\begin{corollary}\label{cor:one-shot averaged}
Let $W_i:\,\X\to\S(\hil_B)$ be channels for $i=1,\ldots,r$, and let 
$\{\gamma_i\}_{i=1}^r$ be a probability distribution. For every $0<\ep<\ep''<1$ and every $\alpha\in(1,2]$,
\begin{align*}
C^{(1)}_{\eps}\bz \sum_i\gamma_i W_i\jz
\le
\max_{i}\chi^*_{\alpha}(W_i)+\frac{1}{\alpha - 1} \log \frac{2}{(1-\ep'')^2}
-\log(\ep''-\ep).
\end{align*}
\end{corollary}
\begin{proof}
Immediate from the second inequality in \eqref{keybounds} and Lemmas \ref{lemma:max cap quasi}
and \ref{lemma:ep-max alpha capacity bound}.
\end{proof}

To prove the lower bound in \reff{keybounds} we will need the following lemma
from 
\cite{HN}:
\begin{lemma}\label{LEMMA_HN}
Consider any channel $W : \cX \mapsto \cD(\cH_B)$. For any $\lambda>0 $, $M \in \bbN$, $p\in\P_f(\X)$ and $c>0$, there exists a code $\cC=(M,\varphi,\Pi)$ such that
\[
p_e(\cC, W)\leq (1+c)\left(1-\sum_x p_x \tr[\{W(x)> \lambda W(p)\}W(x)]\right)+(2+c+c^{-1})\frac{M}{\lambda},
\]
where 
$W(p):=\sum_x p(x)W(x)$.
\end{lemma}

The following Proposition yields the lower bound in \eqref{keybounds}:

\begin{proposition}\label{prop:lower bound}
In the setting of Theorem \ref{thm:one-shot bounds}, we have, for any $p\in\P_f(\X)$,
\begin{equation}\label{relentr lower bound}
C^{(1)}_{\eps}(W)
\ge
D_{\max}^{\sqrt{1-(\ep')^2}}(\rho_{\X B}(p)\|\rho_{\X}(p)\otimes\rho_{B}(p))
+\log\frac{\ep'(\ep-\ep')^2}{8\ep}.
\end{equation}
\end{proposition}
\begin{proof}
Let $0<\ep'<\ep<1$, let $p\in\P_f(\X)$, and $\rho_{\X B}:=\rho_{\X B}(p)$ as in \eqref{rho XB}.
To prove the inequality in \eqref{relentr lower bound}, it is sufficient to prove that there exists a code $\C=(M,\varphi,\Pi)$ 
such that
\begin{equation}\label{lower bound1}
\log M\ge
D+\log\frac{\ep'(\ep-\ep')^2}{8\ep},
\ds\ds\ds D:=D_{\max}^{\sqrt{1-(\ep')^2}}(\rho_{\X B}\|\rho_{\X}\otimes\rho_{B}),
\end{equation}
and $p_e(\C, W)\le \eps$.
Note that if the lower bound in \eqref{lower bound1} is negative then there is nothing 
to prove, and hence for the rest we assume the contrary.

Let $\lambda$ be such that 
$1-\ep'=\Tr\Delta_+(\lambda)$, where
$ \Delta_+(\lambda):= \bigl(\rho_{\X B} - \lambda \rho_{\X} \otimes \rho_B)_+$.
Then
\begin{equation}\label{use}
1-\ep' = \tr  \Delta_+(\lambda) \le \tr[ \{\rho_{\X B} > \lambda\rho_X \otimes \rho_B\} \rho_{\X B} ]=\sum_x p_x \tr[\{W(x)>\lambda W(p)\}W(x)].
\end{equation}
Moreover, Lemma \ref{lemma:smoothing-bound} yields that 
\begin{align}\label{lb proof 2}
D=D^{g(1-\ep')}_{\max}(\rho_{\X B} \| \rho_{\X} \otimes \rho_B)
\le 
\log\lambda-\log\sqrt{1-g(1-\ep')^2}
=
\log\lambda-\log\ep'.
\end{align}

By Lemma~\ref{LEMMA_HN}, for any $c>0$ and $M\in\bN$, there exists a code $\C$ 
of size $M$ such that
\begin{align*}
p_e(\C, W)&\le
(1+c)\left(1-\sum_x p_x \tr[\{W(x)>\lambda W(p)\}W(x)]
\right)+\frac{(1+c)^2}{c}\frac{M}{\lambda}\\
&\le
(1+c)\ep'+\frac{(1+c)^2}{c}M\frac{2^{-D}}{\ep'},
\end{align*}
where the second inequality follows from the choice of $\lambda$.
Such a code surely satisfies $p_e(\C, W)\le \ep$ if the RHS above is upper bounded by $\ep$, or equivalently,
\begin{align*}
M\le \frac{c}{(1+c)^2}2^D\ep\ep'-\frac{c}{1+c}2^D(\ep')^2.
\end{align*}
The RHS of the above inequality is maximal if $c=\frac{\ep-\ep'}{\ep+\ep'}$, which yields the bound
\begin{align*}
M\le 2^D\frac{\ep'(\ep-\ep')^2}{4\ep}=:M_{\ep}.
\end{align*}
Hence,
\begin{align*}
C^{(1)}_{\eps}(W)
\ge
\log\lfloor M_\ep\rfloor
\ge \log M_{\ep}-1=D+\log\frac{\ep'(\ep-\ep')^2}{8\ep}.
\end{align*}
\end{proof}

\noindent\textbf{Proof of Theorem \ref{thm:one-shot bounds}}:
The lower bound in \eqref{keybounds} follows immediately \eqref{relentr lower bound}, by taking the supremum over $p\in\P(\X)$.

To prove the upper bound in \eqref{keybounds}, fix $0<\eps<\ep'' < 1$, and define
\begin{equation*}
\gamma:=\chi^*_{\max,1-\ep''}(W)=\sup_{p\in\P_f(\X)}\chi_{\max,1-\ep''}(W,p).
\end{equation*}
We need to prove that if $\C=(M,\varphi,\Pi)$ is a code such that 
$\log M> \gamma-\log(\ep''-\ep)$ then $p_e(\C, W)>\eps$.

Thus, let $\C=(M,\varphi,\Pi)$ be a code with $\log M> \gamma-\log(\ep''-\ep)$;
then, there exists a $c>1$ such that 
\begin{equation*}
\frac{2^\gamma}{M}=\frac{\ep''-\ep}{c}.
\end{equation*}
Let 
$x_k=\vfi(k),\,k=1,\ldots,M$ be the codewords, and let 
$\rho_k=W(x_k)$ be the output states of the channel.
Let $p\in\P_f(\X)$ be the uniform distribution on the codewords, i.e., $p(x)=1/M$ if $x=x_k$ for some $k=1,\ldots,M$, and $p(x)=0$ otherwise.
For this $p$, we have 
\begin{equation*}
\rho_{\X B}:=\rho_{\X B}(p)=\frac{1}{M}\sum_{k=1}^M\pr{x_k} \otimes \rho_k.
\end{equation*}
Let $0<\delta<\log c$. By the definition of $\chi_{\max,1-\ep''}(W,p)$, there exist
$\bar\sigma_B\in\D(\hil_B)$ and $\bar\rho_{\X B}\in B_{1-\ep''}(\rho_{\X B})$ such that 
\begin{align*}
D_{\rm{max}}(\bar\rho_{\X B}\|\rho_{\X}\otimes\osigma_B)
\le
\chi_{\max,1-\ep''}(W,p)+\delta\le \gamma+\delta.
\end{align*}
Using the definition of $\bar\rho_{\X B}$ and \eqref{fuchs}, we have 
\begin{align}\label{inequality2}
\half\norm{\bar\rho_{\X B}-\rho_{\X B}}_1\le\d(\bar\rho_{\X B},\rho_{\X B})\le 1-\ep'',
\end{align}
and
\begin{align}\label{inequality}
\bar\rho_{\X B}\le 2^{D_{\rm{max}}(\bar\rho_{\X B}\|\rho_{\X}\otimes\osigma_B)}
\left(\rho_{\X}\otimes\bar\sigma_B\right)\le 2^{\gamma+\delta}\rho_{\X}\otimes\bar\sigma_B
=
\frac{2^{\gamma+\delta}}{M}\sum_{k=1}^M\pr{x_k}\otimes\bar\sigma_B.
\end{align}

Let $\hat\Pi:=\sum_{k=1}^M\pr{x_k}\otimes \Pi_k$, which is a projection on $\hil_{\X B}$. Then
\begin{align*}
1-p_e(\C, W)&=
\frac{1}{M}\sum_{k=1}^M\Tr(\rho_k\Pi_{k})
=
\Tr\rho_{\X B}\hat\Pi
=
\Tr(\rho_{\X B}-\bar\rho_{\X B})\hat\Pi
+\Tr\bar\rho_{\X B}\hat\Pi\\
&\le
\half\norm{\bar\rho_{\X B}-\rho_{\X B}}_1+\frac{2^{\gamma+\delta}}{M}\sum_{k=1}^M\Tr\Pi_k\bar\sigma_B\\
&\le
1-\ep''+\frac{2^{\gamma+\delta}}{M}
<
1-\ep''+\ep''-\ep=1-\ep,
\end{align*}
where the first inequality follows from \eqref{inequality}, the second from \eqref{inequality2}, and the last one from the initial assumption on $M$ and the choice of $\delta$.
\hfill$\blacksquare$

\section{From one-shot to asymptotics}\label{sec:asymptotics}

In the asymptotic scenario, one considers a sequence of channels
$\mathbf{W} := \{W^{(n)}\}_{n\in \bN}$, where 
$W^{(n)}:\,\X^{(n)}\to\S(\hil_B^{(n)})$.
A code $\C\n=(M\n, \varphi\n, \Pi\n)$ for $W\n$ and its average error probability $p_e(\C\n,W\n)$ are defined the same way as before, i.e., $M\n$ is a natural number, $\vfi\n:\,\{1,\ldots,M\n\}\to\X^n$ is the encoding map, 
$\Pi^{(n)}:=\{\Pi_i^{(n)}\}_{i=1}^{M_n}$ is the decoding POVM, with each
$\Pi_i^{(n)} \in \B(\hil_B^{\otimes n})$,
and 
\begin{equation*}
p_e(\C\n,W\n)=\frac{1}{M\n}\sum_{i=1}^{M\n}\left[1-\Tr W\n(\vfi\n(i))\Pi^{(n)}_i\right].
\end{equation*}

If there exists 
a sequence of codes $\{\cC^{(n)}\}_{n=1}^\infty$
for which the average probability of error $p_e(\cC^{(n)},W\n) \to 0$ as $n \to \infty$, then {$R:=\liminf_n\frac{1}{n}\log |\C\n|$} is said to be an \ki{achievable rate}. The {\em{(direct) capacity}} $C({\mathbf{W}})$ of the sequence of channels ${\mathbf{W}}$ is defined as the supremum of all achievable rates. 
The corresponding {\em{strong converse {capacity}}} $C^*({\mathbf{W}})$ is defined as the infimum of 
$R$ such that for any sequence of codes $\{\cC^{(n)}\}_{n=1}^\infty$
{with rate $\liminf_n\frac{1}{n}\log |\C\n|\ge R$, we have}
$p_e(\cC^{(n)}) \to 1$ as $n \to \infty$.
It is obvious that 
\begin{equation}\label{capacity inequality}
C({\mathbf{W}})\le C^*({\mathbf{W}}).
\end{equation}
The channel 
is said to satisfy the {\em{strong converse property}} if $C({\mathbf{W}})=C^*({\mathbf{W}})$.

\subsection{Memoryless channels}
\label{sec:memoryless}

We say that ${\mathbf{W}}$ is \textit{memoryless} 
if for every $n\in\bN$,
$\X^{(n)}=\X^n:=\times_{k=1}^n\X$, $\hil_B^{(n)}=\hil_B^{\otimes n}$, and
\begin{equation}\label{memoryless}
W\n(x_1,\ldots,x_n)= W^{\otimes n}(x_1,\ldots,x_n):=W(x_1)\otimes\ldots \otimes W(x_n)
\end{equation}
for any sequence $(x_1,\ldots,x_n)\in\X^n$,
where for simplicity we denote $W^{(1)}$ by $W$.

\begin{remark}
Note that if $W$ is a usual quantum channel, ie., a CPTP map from $\S(\hil_A)$ to $\S(\hil_B)$ then the memoryless extensions $W^{\otimes n}$, defined in \eqref{memoryless}, 
are different from the usual tensor product extensions of $W$. Indeed, one can easily see that our definition of 
$W^{\otimes n}$
coincides with the $n$th tensor product extension of $W$
{\em{with the restriction}} that
only product-state codewords are allowed at the input of the channel. 
Hence, in this case the above defined direct capacity (strong converse capacity) is the so-called {\em product-state} classical capacity (strong converse classical capacity) of the channel. 

Note also that the usual tensor product extension of $W$ is nothing else but the unique factorization of the $n$-linear map given in \eqref{memoryless} through $\S(\hil_A)^{\otimes n}$ (note that $\X=\S(\hil_A)$ in this case).
\end{remark}

For a memoryless channel ${\mathbf{W}}$, we denote the capacity and the strong converse capacity simply as $C(W)$ and $C^*(W)$ respectively, since in this case the sequence of channels, ${\mathbf{W}}$, is given solely in terms of $W$. The capacity of such a channel is given by its Holevo capacity 
$\chi^*(W)$ \cite{Holevo,SW}, and it satisfies the strong converse property \cite{ON,Winter}, i.e., 
\begin{equation}\label{memoryless capacity}
C(W)=C^*(W)=\chi^*(W).
\end{equation}
Here we show how the above identity can be obtained from our 
one-shot bounds in Theorem \ref{thm:one-shot bounds}.

Let ${\mathbf{W}}$ be a memoryless channel. 
By \eqref{capacity inequality}, it is sufficient to show that 
\begin{equation}\label{trivial capacity inequalities}
C^*(W)\le\chi^*(W)\ds\ds\ds\text{and}\ds\ds\ds
C(W)\ge\chi^*(W).
\end{equation}

Note that 
the $\alpha$-capacities are weakly additive, in the sense that \cite{ON}
\begin{equation}\label{additivity}
\chi_\alpha^*(W^{\otimes n})=n\chi_\alpha^*(W),\ds\ds\ds n\in\bN.
\end{equation}
Hence, by Corollary \ref{cor:alpha upper bound}, we have
\begin{equation*}
C^{(1)}_{\eps}(W^{\otimes n})\le n\chi^*_{\alpha}(W)+\frac{1}{\alpha - 1} \log \frac{2}{(1-\ep'')^2}
-\log(\ep''-\ep).
\end{equation*}
for every $0<\eps<\ep''<1$ and $\alpha\in(1,2]$, and $n\in\bN$.
It is easy to verify that 
\begin{align}\label{as cap expression}
C^*(W)&=\lim_{\ep\to 1}\limsup_{n\to\infty}\frac{1}{n}C_{\ep}^{(1)}(W^{\otimes n}),
\end{align}
and hence we obtain
\begin{align*}
C^*(W)&\le\lim_{\ep\to 1}\chi^*_{\alpha}(W)=\chi^*_{\alpha}(W).
\end{align*}
Finally, taking the limit $\lim_{\alpha\searrow 1}$ and using \eqref{alpha cap limit}, we obtain
\begin{align*}
C^*(W)\le\chi^*(W).
\end{align*}

To show the second inequality in \eqref{trivial capacity inequalities}, we first note that 
\begin{align*}
C(W)&=\lim_{\ep\to 0}\liminf_{n\to\infty}\frac{1}{n}C_{\ep}^{(1)}(W^{\otimes n}).
\end{align*}
Let $p\in\P_f(\X)$, and for every $n\in\bN$, let 
$p^{\otimes n}\in\P_f(\X^n)$ be the $n$th 
i.i.d.~extension of $p$, given by $p^{\otimes n}(x_1,\ldots,x_n)=p(x_1)\cdot\ldots\cdot p(x_n),\,x_1,\ldots,x_n\in\X$. 
One can easily see that 
\begin{equation*}
\rho_{\X^n B^n}(p^{\otimes n})=\rho_{\X B}(p)^{\otimes n},
\end{equation*}
and the lower bound in Theorem \ref{thm:one-shot bounds} yields that
\begin{equation*}
C^{(1)}_{\eps}(W^{\otimes n})\ge 
D_{\max}^{\sqrt{1-(\ep')^2}}(\rho_{\X B}(p)^{\otimes n}\|\rho_{\X}(p)^{\otimes n}\otimes\rho_{B}(p)^{\otimes n})
+\log\frac{\ep'(\ep-\ep')^2}{8\ep}
\end{equation*}
for every $0<\ep'<\ep<1$.
Hence, we have
\begin{equation*}
\liminf_{n\to\infty}\frac{1}{n} C^{(1)}_{\eps}(W^{\otimes n})\ge 
\liminf_{n\to\infty}\frac{1}{n} D_{\max}^{\sqrt{1-(\ep')^2}}(\rho_{\X B}(p)^{\otimes n}\|\rho_{\X}(p)^{\otimes n}\otimes\rho_{B}(p)^{\otimes n})
=
D(\rho_{\X B}(p)\|\rho_{\X}(p)\otimes\rho_B(p)),
\end{equation*}
where we used \eqref{strong AEP} for the last identity. Taking the supremum over $p\in\P_f(\X)$ then yields
\begin{equation}\label{asymptotic capacity lower bound}
C(W)\ge\chi^*(W).
\end{equation}

\begin{remark}
Using the standard block coding argument, \eqref{asymptotic capacity lower bound} yields immediately that the 
classical capacity of a memoryless quantum channel (without the product-state restriction) is lower bounded by the regularized Holevo capacity, as in the Holevo-Schumacher-Westmoreland theorem \cite{Holevo,SW}.
\end{remark}

\subsection{Averaged channels}\label{sec:averaged}

We consider a class of channels which are 
convex combinations of a finite number of memoryless channels. 
For a channel in this class, $n$ successive uses is given by the map
$W^{(n)}:\cX^n  \to \cD(\cH_B^{\otimes n})$, defined as
\begin{equation}\label{memory_channel}
W\n=\sum_{i=1}^K\gamma_i W_i^{\otimes n},
\end{equation}
where $\{\gamma_i\}_{i=1}^K$ is a probability distribution (we assume that all the $\gamma_i$ are strictly positive), and for each $W_i:\cX \to\cD(\cH_B)$, $W_i^{\otimes n}$ is the memoryless extension defined in \eqref{memoryless},
i.e., $W_i^{\otimes n}(x_1,\ldots,x_n)=W_i(x_1)\otimes\ldots\otimes W_i(x_n),\,x_j\in\X,\,j=1,\ldots,n$ and $n\in\bN$.
This model describes a scenario in which Alice and Bob know that they are communicating through 
a memoryless channel, but instead of knowing the exact identity of this channel (as in the 
memoryless case), they only know that they are using the channel $W_i$ with probability 
$\gamma_i$.
%
Note that if the first input 
is sent through the channel $W_i$ then all successive 
inputs 
are also sent through {the same channel}. Hence the channel has long-term memory. It is an analogue of the classical averaged channel first introduced by Jacobs \cite{jacobs}. 
Let $C({\mathbf{W}})$ and $C^*({\mathbf{W}})$ denote the capacity and strong converse capacity of the sequence of channels 
${\mathbf{W}} := \{W^{(n)}\}_{n\in \bN}$, respectively.

This long-term memory channel was introduced in \cite{DD}, {where
the authors evaluated $C({\mathbf{W}})$ as
\begin{equation}
C({\mathbf{W}})=\sup_{p\in\P_f(\X)} \min_{1\leq i\leq K} \chi(W_i,p).
\end{equation}
This result was later generalized to more general forms of averaged channels in \cite{BB}.}

Using the fact that the error probability is an affine function of the channel, it can be seen that the strong converse capacity of an averaged channel ${\mathbf{W}}$ is given by 
\begin{equation*}
C^*({\mathbf{W}})=\max_{1\le i\le K}C^*(W_i)=\sup_{p\in\P_f(\X)} \max_{1\leq i\leq K} \chi(W_i,p),
\end{equation*}
where the second identity follows from the memoryless case.
Below we show how the one-shot upper bound of Theorem \ref{thm:one-shot bounds} yields an upper bound on the one-shot capacity of an averaged channel, which in turn yields the inequality
$C^*({\mathbf{W}})\le\max_{1\le i\le K}C^*(W_i)$. For completeness, we give a proof for the converse inequality, too.

Applying Corollary \ref{cor:one-shot averaged} to $W\n=\sum_{i=1}^K\gamma_i W_i^{\otimes n}$, we obtain
\begin{align*}
C^{(1)}_{\eps}\bz \sum_i\gamma_i W_i\jz
\le
n\max_{1\le i\le K}\chi^*_{\alpha}(W_i)+\frac{1}{\alpha - 1} \log \frac{2}{(1-\ep'')^2}
-\log(\ep''-\ep)
\end{align*}
for any $0<\ep<\ep''<1$,
where we have used the additivity of the $\alpha$-capacities \eqref{additivity}. By the same argument as in Section \ref{sec:memoryless}, we obtain that
\begin{align*}
C^*(\mathbf{W})=\lim_{\ep\to 1}\limsup_{n\to\infty}\frac{1}{n}C_{\ep}^{(1)}\bz W\n\jz\le
\lim_{\alpha\to 1} \max_{1\le i\le K}\chi^*_{\alpha}(W_i)= \max_{1\le i\le K}\chi^*(W_i).
\end{align*}

To show that $C^*({\mathbf{W}}) \ge \max_{1\le i\le K} \chi^*(W_i)$,
it suffices to prove that for any $0\le R<\max_{1\le i\le K} \chi^*(W_i)$, there exists a 
sequence of codes $\{\cC_n\}_{n=1}^\infty$ with rate at least $R$ such that
\be\label{not1}
p_e(\cC_n,W^{(n)}) \not\to 1 \quad {\hbox{as}} \quad n \to \infty.
\ee

Thus, let $R$ be as above, and let $j$ be such that
\begin{equation*}
\chi^*(W_j) = \max_{1\le i\le K} \chi^*(W_i).
\end{equation*}
Then it follows from the HSW theorem (\cite{Holevo, SW}; see also \cite{HN}) that
there exists a sequence of codes $\cC^{(n)}$ such that
$\liminf_{n\to\infty}\frac{1}{n}\log|\C_n|\ge R$ and 
\begin{equation*}
\lim_{n\to\infty}p_e(\cC^{(n)},W_j^{\otimes n})
= 0.
\end{equation*}
Hence, if and Alice and Bob use this code to communicate over the long-term memory channel $\mathbf{W}$
then 
\begin{align*}
\limsup_{n\to\infty}p_e(\C\n,W\n)&=
\limsup_{n\to\infty}\sum_{i=1}^n \gamma_i p_e(\C\n,W_i^{\otimes n})\le 1-\gamma_j,
\end{align*}
and the statement follows.

\section{Discussion}
\label{Sec_Discussion}
We have given bounds on the optimal type II error of Stein's lemma in terms of the smoothed max-relative entropy of the two states, and on the one-shot capacity of a channel with error threshold in terms of a quantity analogous to the Holevo capacity, defined again using the smoothed max-relative entropy. 
The smoothed max-relative entropy is a central notion in the
so-called one-shot information theory, which has been a very active and quickly evolving research field in the past few years. The aim of this section is to relate and compare our results to existing results in the field. 

First, a few comments about the 
choice of the distance measure for smoothing. In the original definition of the smoothed min-entropy \cite{RennerPhD}, smoothing was defined with respect to the 
variational distance $d_v$ (half the trace distance), which was replaced in much of the recent works with the so-called purified distance $d_p$ \cite{TCR2}, defined as
\begin{equation*}
d_p(\rho,\sigma):=\sqrt{1-[F(\rho,\sigma)+\sqrt{(1-\Tr\rho)(1-\Tr\sigma)}]^2}
\end{equation*}
for subnormalized states $\rho,\sigma$. 
 In fact, for the type of bounds we considered here, it is quite irrelevant what distance $d$ is used for the smoothing, as long as it is equivalent to the variational distance (in the sense that there 
exist strictly monotone functions $f,g:\,[0,+\infty)\to[0,+\infty)$ such that $g(0)=0$ and $f(d_v(\rho,\sigma))\le d(\rho,\sigma)\le g(d_v(\rho,\sigma))$ for every subnormalized states
$\rho$ and $\sigma$). Indeed, while the concrete form of the smoothing parameter as a function of the error threshold, as well as the form of the additive constants (e.g., in Theorem \ref{prop:hypotesting bounds}), may be different for different distance measures, 
these differences disappear in the asymptotic limit as long the distances are equivalent. In particular, the variational distance, the purified distance $d_p$, the extension $\d$ of the sine distance used in this paper, and the Bures distance
$d_B(\rho,\sigma):=\min_{\{\psi_\rho,\psi_\sigma\}}\norm{\psi_\rho-\psi_\sigma}=\sqrt{\Tr\rho+\Tr\sigma-F(\rho,\sigma)}$ \cite{Bures} are all equivalent on the set of (subnormalized) states, and hence they result in qualitatively equivalent smoothed entropies. The distances $\d,d_p$ and $d_B$, all derived from the fidelity, also seem equally useful for smoothing dual conditional entropies in the sense of \cite{TCR2}.

There are also differences in the choice of the neighbourhood over which smoothing is performed; the main difference here is optimizing over subnormalized states 
in an $\ep$-neighbourhood of the given state $\rho$, or restricting the optimization to normalized states. Again, the difference between the resulting quantities is irrelevant 
for the asymptotic properties of these quantities. We briefly show this here for our definition $ D^\ep_{\max}(\rho\|\sigma)$ of the 
smoothed max-relative entropy (where optimization is restricted to normalized states and the distance is $\d$) and another
common choice \cite{TCR2}, defined as
\begin{equation*}
\tilde D^\ep_{\max}(\rho\|\sigma):=\inf\{D_{\max}(\bar\rho\|\sigma)\,:\,\bar\rho\ge 0,\,\Tr\bar\rho\le 1;\,d_p(\rho,\bar\rho)\le\ep\}
\end{equation*}
(where optimization is over subnormalized states and the distance is $d_p$).
Indeed, let $\rho$ be a state, and $\hat\rho\in B_\ep(\rho)$, where $B_\ep(\rho)$ is the $\ep$-ball around the state $\rho$ with respect to $\d$. Then $\d(\rho,\hat\rho)=d_p(\rho,\hat\rho)$ and hence 
$\hat\rho\in \tilde B_\ep(\rho)$, which implies $\tilde D^\ep_{\max}(\rho\|\sigma)\le D_{\max}(\hat\rho\|\sigma)$, and optimizing over $\hat\rho\in B_\ep(\rho)$
yields $\tilde D^\ep_{\max}(\rho\|\sigma)\le D^\ep_{\max}(\rho\|\sigma)$. 
On the other hand, 
if $\bar\rho\in\tilde B_{\ep}(\rho)$ then $\ep\ge d_p(\rho,\bar\rho)=\sqrt{1-F(\rho,\bar\rho)^2}$, and hence
$\sqrt{1-\ep^2}\le F(\rho,\bar\rho)\le\sqrt{\Tr\bar\rho}$, where the last inequality is due to the monotonicity of the fidelity under the trace. Let $\hat\rho:=\bar\rho/\Tr\bar\rho$. Then
$F(\rho,\hat\rho)=F(\rho,\bar\rho)/\sqrt{\Tr\bar\rho}\ge F(\rho,\bar\rho)\ge \sqrt{1-\ep^2}$, and hence
$\d(\rho,\hat\rho)=\sqrt{1-F(\rho,\hat\rho)^2}\le\ep$, i.e., $\hat\rho\in B_\ep(\rho)$. Thus,
$D_{\max}^{\ep}(\rho\|\sigma)\le D_{\max}(\hat\rho\|\sigma)=D_{\max}(\bar\rho\|\sigma)-\log\Tr\bar\rho\le D_{\max}(\bar\rho\|\sigma)-\log(1-\ep^2)$. 
Optimizing over $\bar\rho\in\tilde B_\ep(\rho)$ yields $D_{\max}^{\ep}(\rho\|\sigma)\le \tilde D_{\max}^{\ep}(\rho\|\sigma)-\log(1-\ep^2)$.
Hence, we finally have
\begin{align}\label{smoothing comparison}
\tilde D^\ep_{\max}(\rho\|\sigma)&\le D^\ep_{\max}(\rho\|\sigma)\le  
\tilde D_{\max}^{\ep}(\rho\|\sigma)-\log(1-\ep^2),\ds\ds\ds\ep\in(0,1).
\end{align}
In particular,
\begin{align*}
\lim_{\ep\searrow 0}\left|\tilde D^\ep_{\max}(\rho\|\sigma)- D^\ep_{\max}(\rho\|\sigma)\right|=0.
\end{align*}
Our main reason to restrict the optimization to normalized states is that otherwise the smoothed max-relative entropy can be negative; in fact, it is easy to see that 
$\lim_{\ep\nearrow 1}\tilde D^\ep_{\max}(\rho\|\sigma)=-\infty$, while $D^\ep_{\max}(\rho\|\sigma)\ge 0$ for any two states $\rho$ and $\sigma$. Since
the smoothed max-relative entropy is a kind of a statistical divergence, or generalized relative entropy, we prefer to keep it non-negative on pairs of normalized states.

In the first version of this paper \cite{DHB}, we used a different type of smoothing, defined as 
$\widehat D^{\ep}_{\max}(\rho\|\sigma):=\inf\{D_{\max}(\bar\rho\|\sigma)\,:\,\bar\rho\ge 0,\,\Tr\bar\rho\le 1;\,\norm{\rho-\bar\rho}_1\le\ep\}$, and gave the bounds
\begin{align*}
\widehat D^{4\sqrt{\ep}}_{\max}(\rho\|\sigma)
\le
-\log\beta_{1-\ep}(\rho\|\sigma)
\le
\widehat D^{\ep/2}_{\max}(\rho\|\sigma)+\log\frac{2}{\ep}
\end{align*}
on the optimal type II error.
Using similar arguments as above, this yields the bounds
\begin{align*}
D^{\sqrt{4\sqrt{\ep}}}_{\max}(\rho\|\sigma)+\log(1-4\sqrt{\ep})
\le
-\log\beta_{1-\ep}(\rho\|\sigma)
\le
D^{\ep/4}_{\max}(\rho\|\sigma)+\log\frac{2}{\ep}
\end{align*}
and
\begin{align}\label{MH bound}
\tilde D^{\sqrt{8\sqrt{\ep}}}_{\max}(\rho\|\sigma)
\le
-\log\beta_{1-\ep}(\rho\|\sigma)
\le
\tilde D^{\ep/4}_{\max}(\rho\|\sigma)+\log\frac{2}{\ep}
\end{align}
in terms of the alternative smoothed max-relative entropies discussed above. 
Using the quantum Stein's lemma, these yield the $\ep$-independent version of Corollary \ref{cor:smooth limit} for $\widehat D^{\ep}_{\max}(\rho\|\sigma)$ in the range $\ep\in(0,1/16)$, for
$D^{\ep}_{\max}(\rho\|\sigma)$ in the range $(0,1/16)$ and for $\tilde D^{\ep}_{\max}(\rho\|\sigma)$ in the range $(0,1/64)$.
Similar bounds were obtained very recently in \cite{TH}, of the form
\begin{align*}
\tilde D^{\sqrt{\ep}}_{\max}(\rho\|\sigma)-\log \nu(\sigma)+\log\ep
\le
-\log\beta_{1-\ep}(\rho\|\sigma)
\le
\tilde D^{\sqrt{\ep-\delta}}_{\max}(\rho\|\sigma)-3\log\delta+3\log 3+\log(1-\ep+\delta),
\end{align*}
where $\nu(\sigma)$ is the number of different eigenvalues of $\sigma$.
These bounds are valid for all $\ep\in(0,1)$ and $\delta\in(0,\ep)$, and hence the quantum Stein's lemma applied to these bounds yields the 
$\ep$-independent version of Corollary \ref{cor:smooth limit} for $\tilde D^{\ep}_{\max}(\rho\|\sigma)$ in the whole range $\ep\in(0,1)$. Using \eqref{smoothing comparison}, these bounds yield
\begin{align*}
D^{\sqrt{\ep}}_{\max}(\rho\|\sigma)-\log \nu(\sigma)+\log\ep(1-\ep)
\le
-\log\beta_{1-\ep}(\rho\|\sigma)
\le
D^{\sqrt{\ep-\delta}}_{\max}(\rho\|\sigma)-3\log\delta+3\log 3+\log(1-\ep+\delta)
\end{align*}
in terms of the smooth entropies used in this paper. Likewise, our bounds in Theorem \ref{prop:hypotesting bounds} yield, with the help of \eqref{smoothing comparison}, the bounds
\begin{align*}
\tilde D^{g(\ep)}_{\rm{max}}(\rho\|\sigma)
\le
-\log\beta_{1-\ep}(\rho\|\sigma)
\le
\tilde D^{\ep'}_{\max}(\rho\|\sigma)+\log \frac{1}{\ep-\ep'}-\log(1-(\ep')^2),
\end{align*}
where $g(\ep):=\sqrt{\eps(2-\ep)}$, and $0<\ep'<\ep$. Apart from the different smoothing conventions, the difference 
between the bounds of \cite{TH} and our Theorem \ref{prop:hypotesting bounds} stems from the different proof methods;
while the bounds of \cite{TH} were derived using an intermediate quantity, the single-shot quantum information spectrum, we used a more direct approach in proving Theorem \ref{prop:hypotesting bounds}, which results in somewhat simpler expressions.

In Section \ref{sec:one-shot} we derived bounds on the one-shot $\ep$-error capacities of a channel $W$ in terms of its 
$\ep$-max capacities, which in the asymptotics gave that the strong converse capacity of $W$ is equal to its Holevo capacity. We emphasize here again that 
in the case where $W$ is a quantum channel, our definition of the (strong converse) capacity gives the (strong converse) capacity for product state encoding \cite{ON,Winter}. The error bound of \cite{ON} actually gives that the unconstrained strong converse rate for arbitrary (i.e., not necessarily product) encoding cannot exceed the infimum (over $\alpha$) of the regularized $\alpha$-capacities; in particular, when the $\alpha$-capacities are additive in the sense that $\chi_\alpha^*(W^{\otimes n})=n\chi_\alpha^*(W)$ for every $n$ and $\alpha$ close enough to $1$, then the unconstrained strong converse rate is equal to the Holevo capacity. Such additivity results were shown in \cite{KW} for a class of quantum channels, including the qudit depolarizing channels and unital qubit channels, thereby providing the first and so far the only examples for quantum channels with the strong converse property with unconstrained encoding.
The error bound of \cite{ON} automatically yields an upper bound on the one-shot $\ep$-error capacities in terms of the $\alpha$-capacities
with $\alpha>1$ (cf.~Corollary \ref{cor:alpha upper bound}), as was already pointed out in Theorem V.1 of \cite{MH}. 
A counterpart of these bounds, i.e., lower bounds on the 
one-shot $\ep$-error capacities in terms of the $\alpha$-capacities with $\alpha\in(0,1)$, have been obtained in \cite{MD,MH}.

It is well-known that channel coding (for classical information) and hypothesis testing are closely related to each other, and that the direct part of the channel coding theorem
(the Holevo-Schumacher-Westmoreland (HSW) theorem \cite{Holevo,SW}) can be recovered using this relation and the quantum Stein's lemma \cite{ON3,HN}.
Explicit bounds on the one-shot $\ep$-error capacity of a channel $W$ in terms of the optimal type II error for discriminating states of the form $\rho_{\X B}=\sum_x p(x)\pr{x}\otimes W(x)$ (cf.~\eqref{rho XB}) from the product of its marginals, have been given in \cite{Wang:2010ux}, which again yields in the asymptotic limit the HSW theorem, i.e., that the (direct) capacity of $W$ is lower bounded by the Holevo capacity of $W$. 
The upper bound of \cite{Wang:2010ux} has been further improved in \cite{MW}, using state discrimination with restricted measurements, and it has been shown in  \cite{Wang:2010ux} that these bounds yield
\begin{equation*}
C_\ep(\mathbf{W}):=\sup_{\{\C_n\}}\left\{\liminf_{n\to\infty}\frac{1}{n}\log|\C_n|\,:\,\limsup_{n\to\infty} p_e(\C_n,W^{(n)})\le \ep\right\}
\le\frac{\chi^*(W)}{1-\ep}
\end{equation*}
for a sequence of i.i.d.~channels with product encoding.
While this is sufficient to determine the direct capacity with weak converse $(\ep\to 0)$, it is not informative for the strong converse capacity $(\ep\to 1)$. In comparison, our approach yields $C_\ep(\mathbf{W})\le \chi^*(W)$ for every $\ep\in(0,1)$, which in particular gives that the strong converse capacity is upper bounded by the Holevo capacity, as we showed in Section \ref{sec:memoryless}.

Bounds on the one-shot $\ep$-error classical capacity of a quantum channel have been given before in \cite{RR}, in terms of a mixture of smoothed min- and max-relative entropies. While these bounds are suitable to obtain the direct capacity of a memoryless channel (with product encoding), they only provide upper bounds on 
the asymptotic $\ep$-error capacity for $\ep$ up to $1/2$, and hence they cannot be used to obtain the strong converse capacity.

\section{Acknowledgments}

ND would like to thank Igor Bjelakovic for a helpful exchange and for pointing out related results for classical and
quantum compound channels. MM was supported by the Marie Curie International Incoming Fellowship {``QUANTSTAT''}.
MH was supported by the UTS Chancellor's Postdoctoral Research Fellowship. FB acknowledges support from the Swiss National Science Foundation, via the National Centre of Competence in Research QSIT. The research leading to these results has received funding from the European {Community's}
Seventh Framework Programme (FP7/2007-2013) under grant agreement number 213681.

\end{document}